\documentclass[11pt]{article}
\usepackage{setspace}
\usepackage{amssymb,verbatim,epsfig}
\usepackage{amsmath,setspace}
 \usepackage[comma,sort&compress,super]{natbib}
\usepackage{multicol}
\usepackage[usenames,dvipsnames]{color}
\usepackage{url}
\usepackage{amssymb, amsmath, verbatim, epsfig}
\usepackage{setspace}

\newcommand{\eq}[1]{Equation\ (\ref{#1})}

\spacing{1}    
\usepackage[hang,small,bf]{caption}

\title{\bf 
\vspace{-1in}
Law of Urination: all mammals empty their bladders over the same duration
}

\author{Patricia J. Yang$^{1}$, Jonathan C. Pham$^{1}$, Jerome Choo$^{1}$ and David L. Hu$^{1,2*}$ \\
\text{\small{Schools of Mechanical Engineering$^1$ and Biology$^2$}} \\ \text{\small{Georgia Institute of Technology, Atlanta, GA 30332, USA}}}
\date{}

\begin{document}
\maketitle 

\noindent{\bf Corresponding author:} \\
David L. Hu \\
801 Ferst Drive, MRDC 1308, Atlanta, GA 30332-0405 \\
(404)894-0573\\
hu@me.gatech.edu\\


\clearpage


\section*{Abstract}
Many urological studies rely upon animal models such as rats and pigs whose urination physics and correlation to humans are poorly understood.  Here we elucidate the hydrodynamics of urination across five orders of magnitude in animal mass. Using high-speed videography and flow-rate measurement at Zoo Atlanta, we discover the ``Law of Urination,'' which states animals empty their bladders over nearly constant duration of $21 \pm 13$ seconds.  This feat is made possible by larger animals having longer urethras, thus higher gravitational force and flow speed.  Smaller mammals are challenged during urination due to high viscous and surface tension forces that limit their urine to single drops.  Our findings reveal the urethra constitutes as a flow-enhancing device, enabling the urinary system to be scaled up without compromising its function.  This study may help in the diagnosis of urinary problems in animals and in inspiring the design of scalable hydrodynamic systems based on those in nature.

\section*{Significance Statement}

Animals eject liquid into environment for waste elimination, communication, and defense from predators. These diverse systems all rely on the fundamental principles of fluid mechanics, which have allowed us to make predictions about urination across a wide range of mammal sizes.  In this study, we report a mathematical model that clarifies misconceptions in urology and unifies the results from over 50 independent urological and anatomical studies.  The theoretical framework can be extended to study fluid ejection from animals, a topic which has received little attention. 

\par
\vspace{0.1 in}


\section{Introduction}
The practice of medical and veterinary urology often relies on simple non-invasive methods to characterize bladder health.  One of the most easily measured characteristics of a bladder system is its flow rate.  Decreases in flow rate can be attributed to unseen problems with the urinary system.  For example, an expanding prostate in aging male animals can constrict the urethra, decreasing flow rate. Obesity can increase abdominal pressure, causing incontinence.  To study these ailments, animal subjects of a range of sizes are often used.  For example, a study of urination in zero gravity involved a rat suspended on two legs for long periods of time\cite{ortiz1999}.  Despite the ubiquitous use of animals in urology studies, little is understood of how the process of urination changes across species and size.

The bladder serves a number of functions, as reviewed by Bentley \cite{bentley1979}.  In desert animals, the bladder stores water to be retrieved in time of need.  In mammals, the bladder is waterproof, acting as a reservoir that can be emptied at a time of convenience.  This control of urine enables animals to keep their homes sanitary and themselves inconspicuous to predators.  Lastly, stored urine can be used as a defense mechanism, as one knows from picking up small rodents and pets.

There persist several misconceptions in urological hydrodynamics that have important repercussions in the interpretation of healthy bladder function.  For instance, several recent investigators state that flow is generated entirely by bladder pressure.  Consequently, their modeling of the bladder completely neglects gravitational forces\cite{rao2003,walter1993,barnea2001}.  Others, such as Martin, contend urine flow is driven by a combination of both gravity and bladder pressure\cite{martin2009}.  In this study, we show Martin's physical picture is the correct one.  Moreover, we provide quantitative evidence that gravitational force becomes dominant in generating flow as animals increase in size.

In this study, we elucidate the hydrodynamics of urination across animal size.  The theoretical models for this work are given in the Supplement.  In \S 2, we report bladder anatomy and measurements of urination duration and flow rate.  Furthermore, we compare our theoretical predictions to our observations.  In \S 3, we discuss the implications of our work and suggest directions for future research.  In \S 4, we summarize the contributions of our study. 
	

\section{Results} 

We film urination of sixteen animals ({\bf Figure 1(a)-(d)}) and obtained twenty-eight urination videos from YouTube.  {\bf Figure 1(h)} shows the duration of urination across 6 orders of magnitude of animal mass from 0.03 kg to 8000 kg.  Despite this large range, the duration remains nearly constant, $T = 21 \pm 13$ seconds (N=32) for all animals heaver than 3 kg.  This constancy of emptying time is quite a feat upon consideration of the substantial bladders of larger animals.  The urination duration for both jetting and dripping regime are modeled in the Supplement. 

\subsection{Mammalian bladders are isometric}
Our model relies on assumptions of isometry of the urinary system, the property that the urinary system maintains constant proportion for all animals.  In this section, we show indeed bladders are isometric and the driving pressure is constant across animal size. 

We employ previously measured urethra geometries for 69 individuals across 10 species,  reprinted in the Supplement.  {\bf Figure 2(a)} shows the relation between animal mass $M$ and urethra dimensions, length $L$ and diameter $D$.  Trends for $L$ and $D$ are nearly parallel, due to their scaling with similar power law exponents, $M^{0.41}$ and $M^{0.38}$, respectively.  Accordingly, the aspect ratio of the urethra is nearly constant $D/L = 0.06M^{-0.03}$, indicating that for all animals the urethra is 20 times longer than it is wide.  For instance, a rat of 0.3 kg has a urethra of diameter 1 mm and length 2 cm, comparable to a coffee stirrer.  In contrast, an elephant of 5000 kg has a urethra of diameter 5 cm and length 1 m, comparable to a drain beneath a sink.  Although the animals span a weight of 6 orders of magnitude, the aspect ratio of their urethra remains constant, indicating the urethra geometry is conserved throughout evolution.  

From side views in ultrasonic imaging, the urethra appears circular with apparent diameter $D$.  In fact, histology shows the urethra is corrugated\cite{dass2001, praud2003}, as shown in {\bf Figure 1(e)}.  These corrugations decreases its cross sectional diameter from $D$ to $\alpha D$ where $\alpha$ is a shape factor.  We define $\alpha = \frac{1}{D}\sqrt{4A / \pi}$, where area $A$ and apparent diameter $D$ are calculated by image analysis from histological cross sections.  The shape factor $\alpha$ is constant among both species and animal size: we find $\alpha = 0.4 \pm 0.15$ (N = 5) across 4 orders of magnitudes in body weight as in {\bf Figure 2(b)}.  Thus, the effective urethra diameter is reduced to $\alpha = 40\%$ of its apparent diameter.  Furthermore, the cross sectional area is reduced to $\alpha^2 = 0.16$ of the circular area, which is nearly equal to the value of 0.17 from Wheeler\cite{wheeler2012}.

{\bf Figure 2(c)} shows the relation between animal mass and bladder capacity.  The volume of the bladder's capacity is $V = 4.6 M^{0.97}$ mL (N = 9), whose exponent, close to unity, indicates that bladders are quite close to isometric.  The observed bladder volume scaling indicates that 1 kg animal has a bladder of 5 mL, or equivalently, the bladder is 5\% of the animal's weight.  

{\bf Figure 2(d)} shows the relation between body mass and maximum bladder pressure $P_{\mathrm{bladder}}$.  Maximum bladder pressure is difficult to measure in vivo.  Approximate values are given using pressure transducers placed within the bladders of anesthetized animals.  Pressure is then measured when the bladder is filled to capacity by the injection of fluid.  This technique yields a constant bladder pressure across animal size: $P_{\mathrm{bladder}} = 4790 \pm 1207 $ Pa (N = 8).  This bladder pressure is equivalent to the gravitational force of a 49 cm tall column of water, using the conversion 1 cm H$_2$O = 98 Pa.  Thus, for animals with a urethra length of 5 cm, the driving force of bladder pressure is ten times the gravitational head.   The constancy of bladder pressure (4.8 kPa) is consistent with other systems in the body that rely on pressure.  The respiratory systems of animals are also known to operate at a comparable constant pressure of 10 kPa, as demonstrated by Kim and Bush\cite{kim2012}.  

We now apply our model, presented in the Supplement, with the assumption of isometry and constant pressure which enable us to simplify the equations of urine motion substantially.

\subsection{Large animals urinate for constant duration}
	Our predictions for urine duration compare quite favorably with experimental values.  As shown in {\bf Figure 1(h)}, in our experiments,  we observe a near invariance of urination time with body mass, $T \sim M^{0.12}$.  We refer to the observed scaling as the {\it Law of Urination}.  Using our model, we predict:
\begin{equation}
	T = \frac{4V}{\pi \alpha^2 D^2 \sqrt{2P_{\mathrm{bladder}} / \rho + 2 g L}}
\label{duration_large}
\end{equation}
	 where $\rho$ is the density of urine and $g$ is gravity.  For large body masses,  \eq{duration_large} has an equivalent scaling  of $T \sim M^{0.16}$, shown by the blue trend line in {\bf Figure 1(h)}.
	  The agreement between the predicted and measured scaling exponents is quite good (0.12 compared to 0.16).  However, numerical values for urination duration are under-predicted by 50 \%. 
	
	How can bladders of both 0.5 kg and 100 kg be emptied in nearly the same duration?  Larger animals have longer urethras, and so greater gravitational force driving flow.  These long urethras increase the flow rate of  larger animals, enabling them to perform the feat of emptying their substantial bladders over approximately the same duration. 

Our model provides a consistent physical picture upon consideration of flow rate.  The flow rate is predicted to scale as $Q \sim M^{2/3}(2P_{\mathrm{bladder}} / \rho + M^{1/3} )^{1/2}$.  If gravitational forces are dominant over bladder pressure, we expect a flow rate $Q \sim M^{5/6} \sim M^{0.83}$.  {\bf Figure 2(e)} shows the relation between animal mass and flow rate.  Our prediction, given by the black line, is comparable to the flow rate for both females, for which $Q_F \sim M^{0.66}$, and for males, for which $Q_M \sim M^{0.92}$.  Both measured exponents are within 20\% of the predicted exponent, indicating our modeling captures the essential physics of the flow.   

\subsection{Small animals urinate quickly and for constant duration}
	{\bf Figure 1(a)-(d)} shows discrete urination styles according to body size: animals larger than 3 kg produce jets and sheets, whereas animals lighter than 1 kg produce only drops.  In this section, we focus on this dripping regime.  For rats and bats, and possibly the juveniles of other species, urination is a high-speed event of 0.01 to 2 second duration, and so quite different from the jetting regime occurring over 20 seconds.
	
	{\bf Figure 1(h)} also shows urination duration is between 0.05 to 2 seconds among the eleven small animals tested, including one bat, five rats and five mice.  The large error bars for the rats are due to varying urine volume across individuals.  
	Our theory in Supplement predicts that animals of small sizes should also urinate for constant duration. Given the anatomy of a rat, we predict the time to fill a drop of urine by a rat is 0.7 s, which is clearly much shorter than for larger animals. 
	
	{\bf Figure 3(a)(b)} and Supporting Video illustrates the urination process for a 0.03 kg lesser dog-faced fruit bat and 0.3 kg rat, both of which release urine drops.  {\bf Figure 3(c)} shows the time course of the radius of the urine drop formed by the rat.  The trend lines denote prediction of our model given in the Supplement, with varying values of the shape factor $\alpha$.  Inputs to the model include bladder pressure, urethra dimensions and shape factor ($P_{\mathrm{bladder}}$ = 6031 Pa, $L$ = 20 mm, $D$ = 0.8 mm).  Without consideration of the corrugated cross section, a prediction of $\alpha = 1$ (green line) yields a flow rate that is far too high, as shown by the green line.  Using the value from literature\cite{praud2003} of $\alpha = 0.5$, our model predicts a flow speed of $u$ = 1.9 m/s, shown by the the red line, which fits the data fairly well.  Using nonlinear least-squares fitting in Matlab, the experimental data is best fitted with a shape factor of $\alpha = 0.4$, shown in blue.  We conclude the actual shape factor is between 0.4 and 0.5, and the dynamics of drop creation are quite sensitive to this value.
	
	Our model does a fair job of predicting the size of the drop falling from rats and mice.  For rats, the final drop radius $R_f$ before detachment is 3.3 mm, and the time required to fill the drop $T_{\mathrm{drop}}$ is 0.15 s; for mice, $R_f$ = 2.2 mm and $T_{\mathrm{drop}}$ = 0.14 s.  These values are all 50\% larger than predicted by our model in the Supplement.  In order for such a large drop to remain attached, we require the attachment diameter to be larger by a factor of 8.  We conclude drops remain attached to the high density of hair surrounding the urethra opening.  Indeed in experiments, we observe drops can remain attached for minutes at a time before release. 
	
	The model gives the insight on the physical constrain of animal to eject droplets.  In our model, the flow speed is positive only if $P_\mathrm{bladder} \alpha D \geq 4 \sigma$, where $\sigma$ is the surface tension of urine, which is comparable to water\cite{ogata1970}.  Thus, we predict the smallest urethra to expel urine is $4 \sigma / \alpha P_\mathrm{bladder} \sim$ 0.15 mm.  According to our allometric trends, this urethra size corresponds to a body mass of 1.3 g and urethra length  of 2.2 mm.  This length corresponds to the body size of insects such as aphids and water treader {\it Mesovelia}, which excrete liquid drops from their anuses\cite{bush2007}, as well as altricial mice which require their mother's assistance to excrete urine drops\cite{moore1986}.
	
\section{Discussion} 
In our model, we applied two assumptions.  First, the bladder is full before the animal urinates.  Second, we assume the urethra is vertical while the animal is urinating, which is true at least for male animals.  According to an unpaired $t$ test, sex did not affect flow duration or speed (two-tailed $p$ value of $p = 0.15$).  Thus, despite a difference in urethra angle, female and male data sets are not statistically distinct.  

The urinary system is remarkably good at preserving function across a wide range of scales.  This robustness is mainly due to the contribution of the urethra, a tube whose hydrodynamic consquence was previously unknown.  In the medical literature, the urethra is simply known as a conduit between bladder and genitals.  In this study, we find the urethra is analogous to Pascal's Barrel, acting as an energy input device.  By providing a water-tight pipe to direct urine downward, the urethra increases the gravitational force acting on urine and so the rate that urine is expelled from the body.   Thus, the urethra is critical to the bladder's ability to empty quickly as the system is scaled up.  

\section{Materials and Methods}
	\subsection{Animal preparation and filming}
	Animals are filmed at a combination of locations, including Zoo Atlanta, Atlanta Humane Society (AHS), Georgia Tech, and the Animal and Dairy Science Department at the University of Georgia (UGA).  The numbers of animals and their location is given in the Supplement in Tables 1-6.  We film urination of animals using high-speed cameras (Vision Research v210 and Miro 4).  The masses of animals are provided by the keeper from annual veterinary procedures, or massed using an analytical balance.	
	
	\subsection{Anatomical measurements}
	Length $L$ and diameter $D$ of urethra are collected from a combination of X-rays and ultrasound data from previous literature.  Bladder volume $V$ are collected from a combination of filling cystometrography and ultrasonography from previous literature.  Bladder pressure $P_\mathrm{bladder}$ is reprinted here from previous literature.  We reported the measurements and the corresponding body masses, listed in Table 2-6 in the Supplement.
	
	\subsection{Flow rate measurement}
	Flow rate $Q$ of five intermediate-sized animals is measured by simultaneous high speed videography and manual urine collection.  Containers of appropriate size and shape are used to collect the urine of animals.  Flow rate of eight animals are obtained from the literature with estimated corresponding body masses, listed in Table 7 in the Supplement.  Flow rate of small animals is estimated using high-speed videography.  Using the open source software Tracker, we measure the growth in radius and eventual release of urine drops produced by rats.

\section*{Acknowledgements} 
We acknowledge our funding sources (NSF CAREER PHY 1255127), photographer C. Hobbs, and hosts at the Atlanta zoo (R. Snyder) and at animal facilities at Georgia Tech (L. O'Farrell).

\clearpage

 \clearpage

      \begin{table*}
           \begin{center}
                \includegraphics[width = 6.5in,keepaspectratio=true]
             {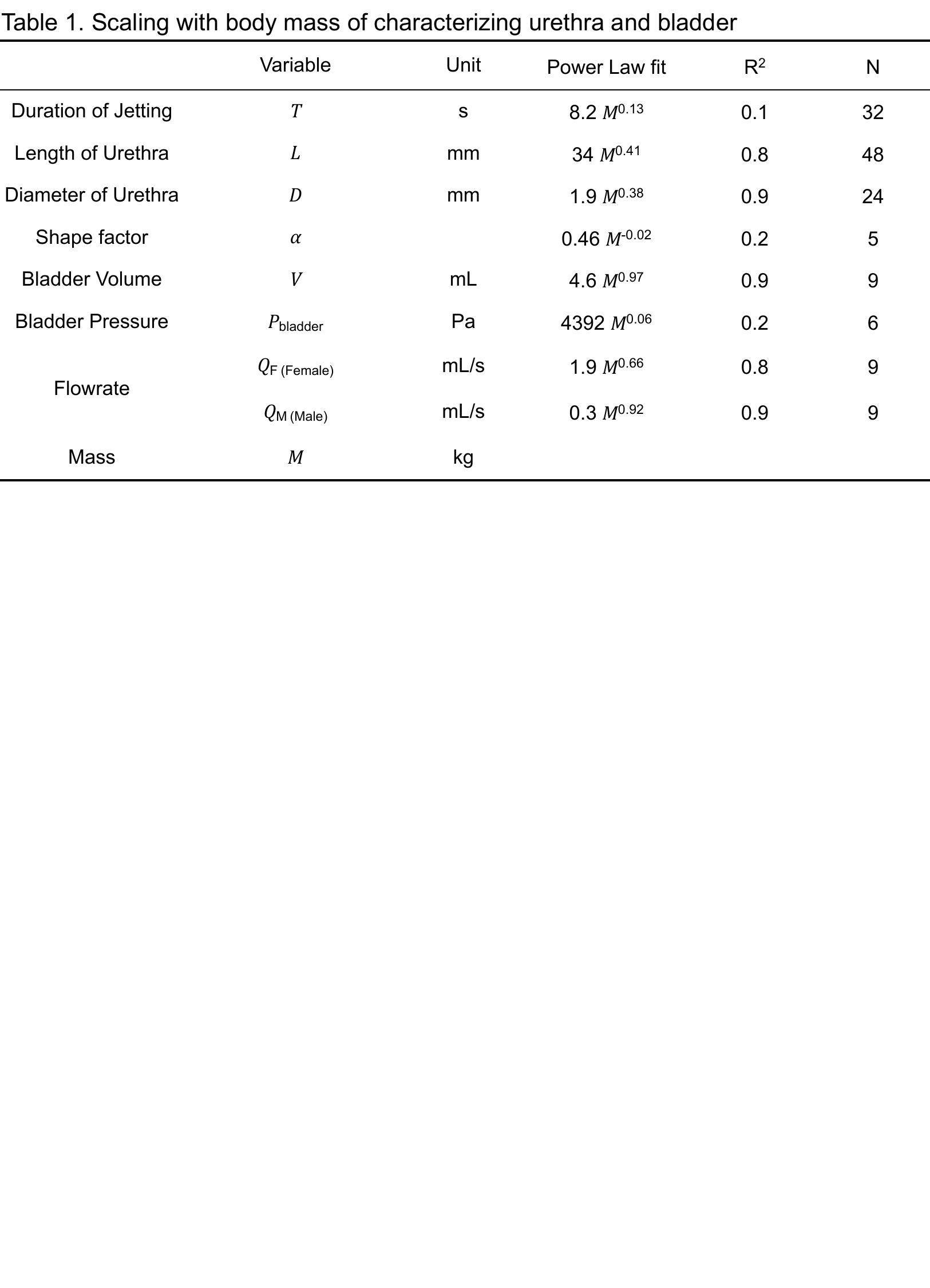}
         \end{center}
    \label{tabletrend}
    \end{table*}
    
     \begin{figure*}
     \begin{center}
             \includegraphics[width = 7in,keepaspectratio=true]
          {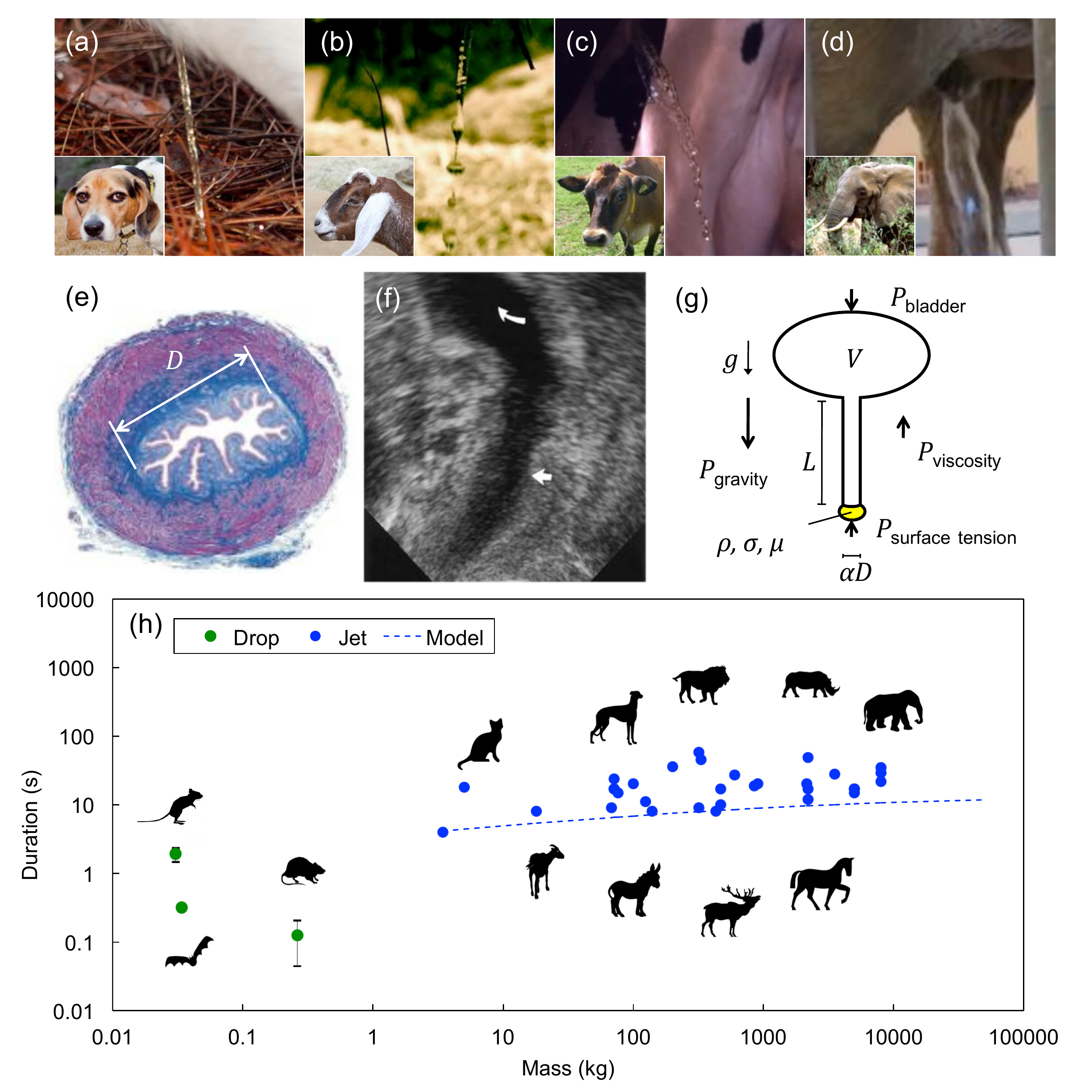}
     \end{center}
    \caption{Jetting urination by large animals, including dog (a), goat (b), cow (c) and elephant (d).   (e) Transverse histological sections of the urethra from a female pig, reprinted from Dass, Greenland and Brading\cite{dass2001}.  (f)  Ultrasound image of bladder and urethra of a female human, reprinted from Prasad\cite{prasad2005}.  The straight arrow indicates the urethra and curved arrow the bladder.  (g) Schematic of the urinary system.  (h) The relation between body mass and urination duration.  Insets of cow and elephant as well as all animal sillhouettes are from public domain\cite{cow, elephant,bat}.}
      \label{timeplot}
  \end{figure*}
     
    \begin{figure*}
       \begin{center}
                \includegraphics[width = 7in,keepaspectratio=true]
             {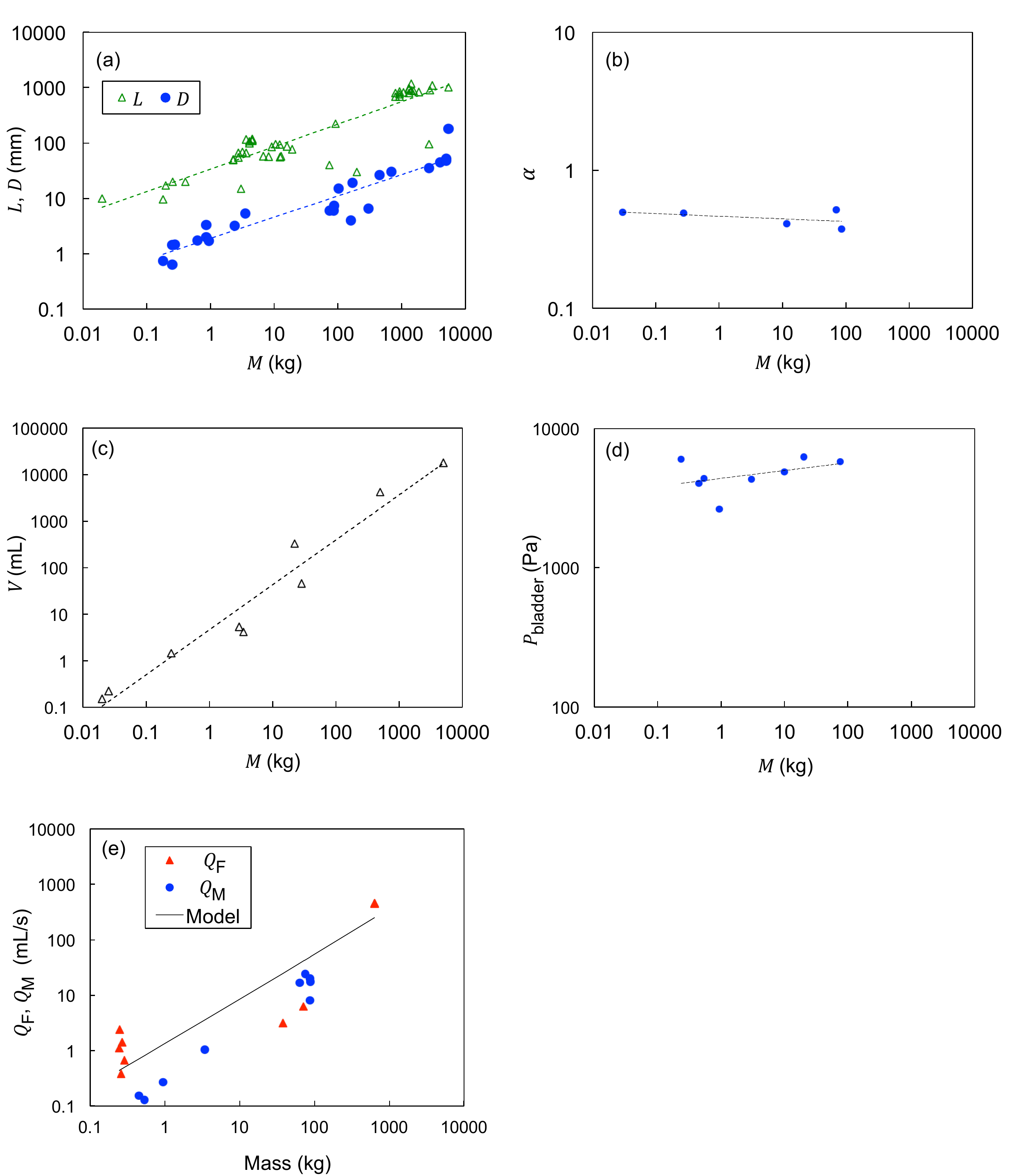}
         \end{center}
   \caption{The relation between body mass $M$ and properties of the urinary system.  (a) Length $L$ (green triangles) and diameter $D$ (blue dots) of urethra.  (b) Shape factor $\alpha$ for cross section of urethra.  (c) Bladder volume $V$.  (d) Bladder pressure $P_{\mathrm{bladder}}$.  (e) Flowrate of urine from experiment and model,  including flow rate $Q_\mathrm{F}$ for female (red triangles), $Q_\mathrm{M}$ for male (blue dots), and the predication of flowrate (black line).  The length of urethra $L$, diameter of urethra $D$, bladder volume $V$ and flow rate $Q$ increase with body size.  The bladder pressure $P_{\mathrm{bladder}}$ and the shape factor $\alpha$ are independent of animal size.}
    \label{trendline}
    \end{figure*}   
 
      \begin{figure*}
       \begin{center}
                \includegraphics[width = 6in,keepaspectratio=true]
             {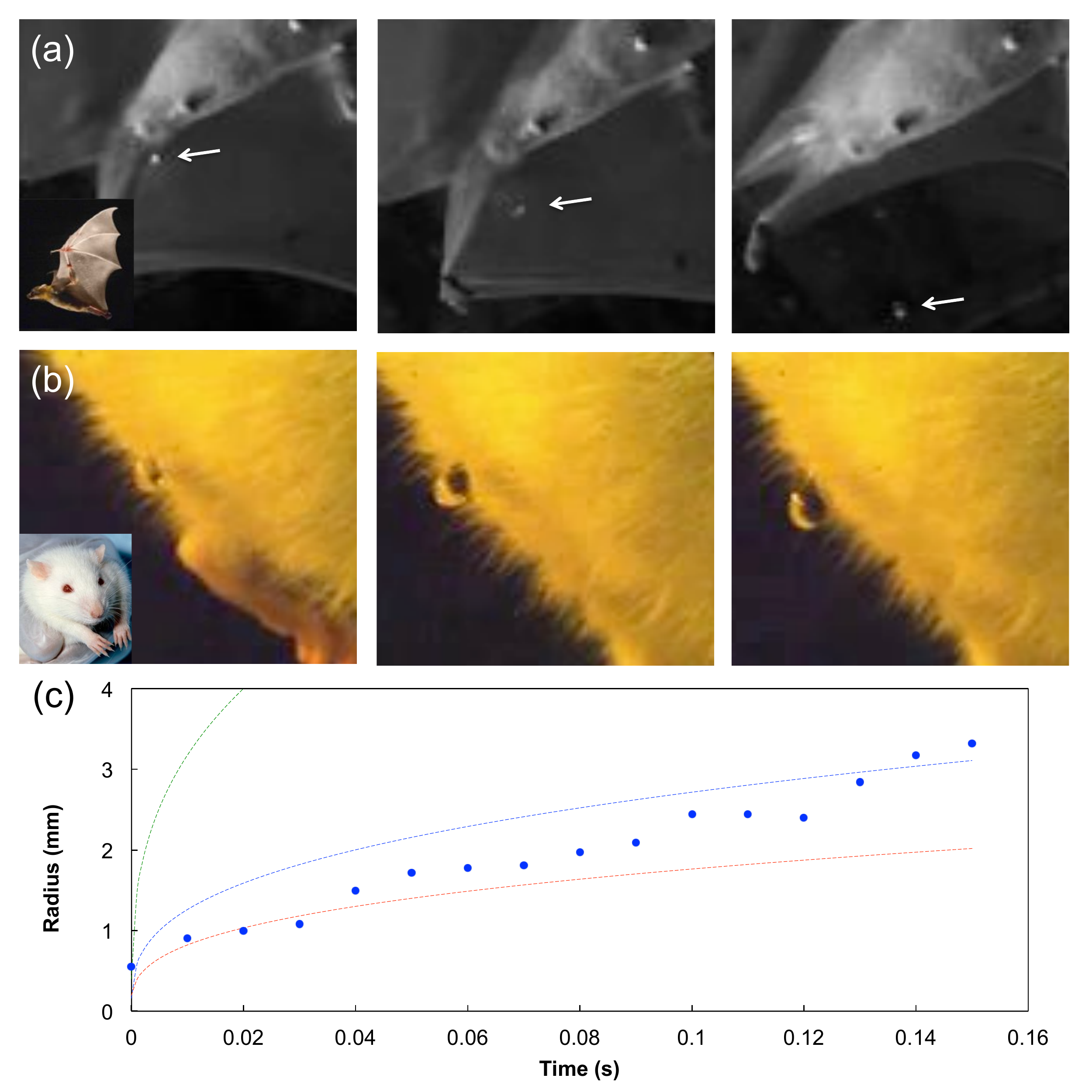}
         \end{center}
      \caption{Dripping urination by small animals.  (a) A lesser dog-faced fruit bat {\it Cynopterus brachyotis} releases a urine drop, reprinted courtesy of Kenny Breuer and Sharon Swartz, Brown University  (b) A rat's urine drop grows with time. (c)  The time course of the drop radius (dots), along with model predictions (lines) using various values of $\alpha$.  Green line represents the prediction with $\alpha = 1$;  Red line is the prediction with $\alpha = 0.5$ from anatomical measurement of rat\cite{praud2003}.  Blue line is the best fit to the experiment with $\alpha = 0.4$.}
      \label{drops}
    \end{figure*} 

\clearpage

\clearpage
\section*{Supporting Information}
{\bf Equations:} Hydrodynamic urination model for large and small animals. \\ 
{\bf Materials and Methods:} Detailed methods on data collections and measurements.\\ 
{\bf Table 1:} Duration of urination. \\
{\bf Table 2:} Length of urethra from literatures. \\
{\bf Table 3:} Diameter of urethra from literatures. \\
{\bf Table 4:} Shape factor of cross section of urethra from literatures.  \\
{\bf Table 5:} Bladder volume from literatures. \\
{\bf Table 6:} Bladder pressure from literatures.  \\
{\bf Table 7:} Flowrate of urination. \\
{\bf Figure 1:} The magnitude of five dimensionless group. \\


\clearpage

 \section*{Hydrodynamic urination model for large and small animal}
 \subsection*{Governing equation for urinary system}
We begin with a number of simplifying assumptions regarding the urinary system.  
The urethra consists of a straight pipe of length $L$ and diameter $\alpha D$.  According to our literature findings, the bladder of volume $V$ holds much more volume than the urethra $\pi \alpha^2 D^2 L / 4$.  We neglect changing height of urine in the bladder.  Urination begins when the smooth muscles of the bladder increase the pressure of its contents to $P_{\mathrm{bladder}}$, measured relative to atmospheric pressure.  After an initial transient, a steady flow of speed $u$ is generated.

We model the urinary flow at steady-state, high Reynolds number and the urine as an incompressible fluid of density $\rho$, viscosity $\mu$ and surface tension $\sigma$.  We apply the energy equation, also known as a modified Bernoulli's Equation, between the entrance and exit of the urethra.  We also assume the urethra is vertical, but discuss modifications of the urethra angle in the Discussion section.  The equation relating the average flow velocity $u$ of the urine to the geometry of urethra: 

\begin{equation}
	P_{\mathrm{bladder}} + P_{\mathrm{gravity}}= P_{\mathrm{inertia}}+ P_{\mathrm{viscosity}}+P_{\mathrm{surface\_ tension}} \ ,
\label{goveringeq}
\end{equation}  

The left hand side of the equation are the two driving pressures, bladder pressure and gravitational pressure associated with a fluid column of height $L$ within the urethra.  These driving pressures eject the urine waste products away from the body.  By conservation of energy, these pressures are transferred to a term describing the kinetic energy of the fluid, $\rho u^2 /  2$, which the animals try to maximize in order to eject flow the fastest.  The remaining terms on the right hand side, viscous losses and surface pressures, represent ``loss'' terms because they resist the flow of urine away from the body.

We formulate the viscous losses using the Hagen-Poiseuille equation, $P_{\mathrm{viscosity}} = 128\mu QL / \pi \alpha^4 D^4$, which describes the pressure drop due to viscosity in a long cylindrical pipe. The average flow rate is $Q = u \pi \alpha^2 D^2 / 4$.  The Hagen-Poiseuille equation maybe be written as $P_{\mathrm{viscosity}} = 32 \mu u L/ \alpha^2 D^2$.

In slow urination, individual drops at the end of the urethra can also resist the flow of urine, similarly to a balloon held at the end of a faucet.  The fluid within these drops has an overpressure given by the Laplace pressure $P_{\mathrm{surface\_ tension}} = 4 \sigma / \alpha D$, where we assume the drops to first order are spherical and the same diameter as the urethra.  Higher-order approximations would take into account the pendular shape of the drop.  We substitute loss terms due to viscosity and surface tension into \eq{goveringeq}, arriving at:

\begin{equation}
P_{\mathrm{bladder}} + \rho gL = \frac{\rho u^2}{2} + \frac{32 \mu uL}{ \alpha^2 D^2} + \frac{4 \sigma}{ \alpha  D}.
\label{forcebalance}
\end{equation}
where $g$ is the gravitational acceleration.  The relative magnitudes of the five pressures enumerated in \eq{forcebalance} are prescribed by five dimensionless groups, the Froude $Fr$, Bond $Bo$, Reynolds $Re$ numbers, aspect ratio $As$ and pressure ratio $Pb$, respectively defined by
\begin{eqnarray*}
Fr &=& \frac{u}{\sqrt{gL}} =  \mathrm{\frac{inertia}{gravity}}
\qquad Bo = \frac{\rho g  D^2}{\sigma} = \mathrm{\frac{gravity}{surface \ tension}} \qquad \\
Re &=& \frac{\rho u D}{\mu} = \mathrm{\frac{inertia}{viscous}}
\qquad As = \frac{D}{L}  = \mathrm{\frac{diameter}{length}} \\
Pb &=& \frac{P_{\mathrm{bladder}}}{\rho gL} = \mathrm{\frac{bladder \  pressure}{gravity}}.
\end{eqnarray*}

The dimensionless energy balance may be written
\begin{equation}
Pb + 1 =  \frac{1}{2} {Fr}^2+ 32 \ \frac {Fr^2} {\alpha^2 As \ Re } + 4 \ \frac{As} {\alpha Bo}.
\end{equation}
In the following sections, we solve this equation in the limit of large and small animal sizes.  

\subsection*{Urination time for intermediate and large animals}
\label{largeanimal}
We define intermediate-sized animals to be those with mass above $M =1$ kg.  {\bf Figure 1} in the Supplement shows the scale of the dimensionless groups in this regime.  Specifically, $Re$ = $10^2-10^4$, $Bo$ = $1-10^2$, $Fr$ = $0.1-1$, $As$ = 0.1, and  $Pb$ = $1-10$.  Thus, large animals have urethras of sufficient diameter that $Bo \gg 1$ and sufficient speed that $Re \gg 1$.   The viscous losses and surface energy of the drop formed at the end of the urethra is negligible compared to the energy associated with the bladder pressure, gravity, and inertia of the fluid.  In this limit, our governing equation, \eq{forcebalance}, reduces to:

\begin{equation}
P_{\mathrm{bladder}} + \rho gL = \frac{\rho u^2}{2}.
\label{largeanimal}
\end{equation}

Assuming $P_{\mathrm{bladder}}$ is constant, as will be shown in the Results section, the steady flow speed may be written in terms of bladder pressure and urethra length, $u = \sqrt{(2P_{\mathrm{bladder}} / \rho+ 2gL)}$.  The urination duration $T$ is the time to completely empty the bladder, $T = V/Q$, where $V$ is the bladder volume.  The duration can be written, 

\begin{equation}
T = \frac{4V}{\pi \alpha^2 D^2 \sqrt{(2P_{\mathrm{bladder}} / \rho+ 2gL)}}.
\label{duration}
\end{equation}  

We apply scaling to predict the trend of duration with animal size.  Since animal bladder systems are isometric, we scale the bladder volume $V \sim M$, length of urethra $L \sim M^{1/3}$ and diameter $D \sim M^{1/3}$.  The shape factor $\alpha$ is constant, as will be shown in the Results section.  Using these scalings we find the flow rate $Q$ is 
\begin{equation}
Q \sim M^{2/3} \left( \frac{2P_{\mathrm{bladder}}}{\rho}  + M^{\frac{1}{3}} \right) ^{\frac{1}{2}}.
\label{Q}
\end{equation}

Thus, we begin to see why large animals can empty their bladders in the same duration as smaller animals.  Larger animals have longer urethras which provide greater gravitational potential energy and so an increase in flow speed.  The combination of a faster flow and wider urethras leads to greater flow rate.  The duration of urination, using \eq{duration} may be written 
\begin{equation}
T \sim M^{\frac{1}{3}} \left( \frac{2P_{\mathrm{bladder}}}{\rho}  + M^{\frac{1}{3}} \right) ^{-\frac{1}{2}}.
\end{equation}
As $M$ increases in size that $M^{1/3} \gg P_{\mathrm{bladder}} / \rho$, the duration $T$ approaches $M^{1/6}$.

\subsection*{Urination time for small animals}
We define small animals as those with $ M < 1$ kg.  Small animals have such thin urethras ($D \leq$ 2 mm) that viscous forces slow the ejection of fluid.  The Supporting Information shows the scale of the dimensionless groups in this regime.  Specifically, $Re = 100 - 3000$, $Bo = 0.1 - 0.5$, $Fr = 0.1 - 6$, $As \approx 0.1$, and $Pb = 10 - 40$.  Thus, small animals have thin short urethra such that the gravitational pressure is negligible compared to the energy associated with bladder pressure, viscosity, surface tension and inertia of the fluid.   In this limit,  \eq{forcebalance} reduces to:

\begin{equation}
P_{\mathrm{bladder}} = \frac{\rho u^2}{2} + \frac{32 \mu u }{As \alpha D} + \frac{4 \sigma}{\alpha D}.
\label{smallanimal}
\end{equation}

We rewrite the equation above by consideration of the variables $P_\mathrm{bladder}, u$ and $D$.  For the intermediate Reynolds numbers for small animals, \eq{smallanimal} can only be solved numerically. However, we can gain insight into the physics of flow for low Reynolds numbers for which the inertia term $\rho u^2/2$ is negligible.  In this regime, the flow speed $u$ is driven by bladder pressure and resisted by viscosity and surface tension.  



For small animals whose urination occur in discrete drops, we begin with the formation of a single drop.  We define radius of drop $R(t)$ as function of time.  The drop is initially the same diameter as urethra $R(0) = \alpha D / 2$.  
The volume of one drop is $V(t)_{\mathrm{drop}} =  4 \pi R(t)^3 / 3$.  The flow rate is the time derivative of $V_{\mathrm{drop}}$ which, by conservation of mass, is equal to flow rate in the urethra $dV_{\mathrm{drop}} / dt = \pi \alpha^2 D^2 u / 4$.  Thus the radius of drop maybe written

\begin{equation}
R(t) = \left( \frac{\alpha^3 D^3}{8} + \frac{3\alpha^2 D^2ut}{16} \right)^{\frac{1}{3}}.
\label{R_final}
\end{equation}

Calculating flow speed $u$ from \eq{smallanimal} and substitute $u$ into \eq{R_final}, we have the radius of drop as function of time.  The drop falls after a duration $T_{\mathrm{drop}}$ passes, when the drop achieves final radius $R_f$.  Using \eq{R_final}, the relationship between duration $T_{\mathrm{drop}}$ and final drop radius $R_{f}$ may be written: 

\begin{equation}
T _{\mathrm{drop}}= \frac{16}{3u\alpha^2 D^2}\left( R_f^3 - \frac{\alpha^3 D^3}{8}\right).
\label{T_drop}
\end{equation}

By consideration of a vertical force balance, we find the final drop radius before it detaches.  The drop falls when its weight $ F_g = 4\pi R_f^3 \rho g /3$, overcomes its attaching surface tension force to the urethra $F_s = \pi \alpha D \sigma \cos \theta$ where $\theta$ is the contact angle of urine on the urethra relative to vertical.  Equating these two forces yields the final drop radius before detachment,
\begin{equation}
R_f^3 = \frac{3 \sigma \alpha D \cos \theta}{4 \rho g}.
\label{R_f}
\end{equation}
Using this relationship for the final drop radius $R_f$, the time to eject one drop may be written 
\begin{equation}
T_{\mathrm{drop}} = \frac{16\alpha D }{3u} \left( \frac{3 \cos \theta }{4\alpha^2 Bo} - \frac{1}{8} \right) \approx \frac{4D\cos \theta}{\alpha Bo} \frac{1}{u}
\label{Tdrop}
\end{equation}


We now use isometry to scale the duration for small animals at low $Re$.  First, the animal is isometric thus $V \sim M$ and $D \sim M^{1/3}$.  Due to the linear scaling of $u$ on $D$, the time to eject one drop from \eq{Tdrop} scales as $T_{\mathrm{drop}} \sim Bo^{-1} \sim M^{-2 / 3}$.  The final drop size from \eq{R_f} is $R_f^3 \sim D \sim M^{1 /3}$.  The full bladder of volume $V$ can produce $N$ drops where, 
\begin{equation}
N = 3V / 4\pi R_f^3 \sim M^{2/3}.
\end{equation}

Thus, the duration $T = N T_{\mathrm{drop}} = $ constant, and so independent of animal size.

\section*{Materials and Methods}
	\subsection*{Animal preparation and filming}
	Animals are filmed at a combination of locations, including Zoo Atlanta, Atlanta Humane Society (AHS), Georgia Tech, and the Animal and Dairy Science Department at the University of Georgia (UGA).  The numbers of animals and their location is given in the Supplement in Tables 1-6.  We film urination of animals using high-speed cameras (Vision Research v210 and Miro 4).  Intermediate-sized animals, such goats urinate naturally during their daily activities.  Dogs urinate as they are released for their regularly scheduled exercise period throughout the day.  Cows are enticed to urinate as part of their normal health monitoring conducted by the staff of UGA, during which urination is elicited by manual simulation of the vulva.  Large animals such as elephants are filmed urinating as soon as they are released to their outdoors pens in the morning.  The masses of elephant, goat and cow are provided by the keeper from annual veterinary procedures.  
	The urination of small animals is a high-speed event and certain preparations increase the chance of capturing it on film.  Before filming, mice and rats are kept in a night-day cycled room for one week.  Water is provided to mice and rats only during the day cycle.  This restriction increases the frequency of daytime drinking, ensuring a full bladder is encountered during our experiments. Mice and rats urinate as a defense response to being removed from their cages.  They are massed using an analytical balance. 
	
	\subsection*{Urination duration measurement} 
	Urination duration is observed by high-speed film.  The onset of urination is defined as the emergence of the first droplet noticeably extruded from the urethra.  The completion of urination occurs when the last drop exits the urethra.   We film one elephant, two goats, five rats and five mice and obtained twenty-nine additional urination videos of various animals from YouTube, listed in Table 1 in the Supplement.   The corresponding masses of the adult animals are found from literature \cite{nowak1999,mattern2000,wilcox1997,wilson2001,miller1997,brown1996,starkey1992,lynette2013,linnaeus1758,kingdon1988,marvin1992,bongianni1988,potts1997,toon2002,shoshani1982}.
	
	\subsection*{Anatomical measurements}
	Length $L$ and diameter $D$ of urethra are collected from previous investigators which use a combination of X-rays and ultrasound to measure urethral geometry.  We have forty-eight measurements of urethral length from literature\cite{clair1999,chang2000,kamo2004,vogel2007,johnston1985,kropp1998,takeda1995,prasad2005,kohler2008,lueders2012,hildebrandt2000,fowler2006} and twenty estimated corresponding body masses\cite{krumrey1968,nowak1999}.  Similarly, twenty-four urethral diameters are obtained from literature\cite{chang2000,de2008,russell1996,kunstyvr1982,root1996,wang2010,gray1918,tsujimoto2003,pozor2002,hildebrandt1998,bailey1975,hildebrandt2000,fowler2006} and twelve corresponding body masses\cite{tasaki2009,perrin2003,sturman1985,scott1970,lein1983,ogden2004,nowak1999}, listed in Table 2 and Table 3 in the Supplement.
	
	We determine the cross section of urethra $A$ by image analysis.  We measure the maximum diameter $D$ and the cross section area $A$ from the anatomical pictures from literature \cite{treuting2011,caceci,praud2003,dass2001,skarva} and two estimated corresponding body masses \cite{wilcox1997,ogden2004}.  The urethra has stared shape instead of circular.  The cross section is deduced by the irregular folded of the urethra.  Thus we define the shape factor $\alpha$ to measure the deduction, which is defined as $A = \pi (\alpha D)^2 /4$, as listed in Table 4 in the Supplement.  For the following analysis, we assume the urethra is a circular tube with the equivalent diameter $\alpha D$.
	
	Bladder volume $V$ are collected from previous investigators which use a combination of filling cystometrography and ultrasonography.  We have eight measurements of bladder volume from literature \cite{pandita2000, birder2002, herrera2010, thor1995, abdel2001, atalan1998, higgins2006, fowler2006}.  Bladder pressure $P_\mathrm{bladder}$ from six animals are reprinted here from previous investigators, who use a pressure transducer to measure the bladder pressure changes with time\cite{ishizuka1996,hinman1971,van1995,schmidt2003}.  We report the maximum value of $P_\mathrm{bladder}$ during the urination cycle.  Values of pressure are given relative to atmosphere pressure.  Bladder volume $V$ and bladder pressure $P_\mathrm{bladder}$ are listed in Table 5 and 6 in the Supplement.
	
	\subsection*{Flow rate measurement}
	Flow rate $Q$ of five intermediate-sized animals is measured by simultaneous high speed videography and manual urine collection.  Containers of appropriate size and shape are used to collect the urine of goats and cows.  Dog urine is collected and weighed using a urination mat for pets; volume is approximated by assuming urine has the density of water, as is known by Ogata\cite{ogata1970}.   The average flow rate is given by the volume of urine divided by the duration of urination.  Flow rate of eight animals are obtained from the literature\cite{van1995,schmidt2003,masumori1996,folkestad2004,tsujimoto2003} with estimated corresponding body masses \cite{schmidt2002,ogden2004}, listed in Table 7 in the Supplement.  Flow rate of small animals is estimated using high-speed videography.  Using the open source software Tracker, we measure the growth in radius and eventual release of urine drops produced by rats.

\clearpage
\bibdata{urinebib}
\bibliographystyle{pnas2009}
\bibliography{urinebib}

\begin{thebibliography}{10}

\bibitem{ortiz1999}
Ortiz, R.~M, Wang, T.~J,  \& Wade, C.~E.
\newblock (1999) Influence of centrifugation and hindlimb suspension on
  testosterone and corticosterone excretion in rats.
\newblock {\em Aviation, Space, and Environmental Medicine} {\bf 70}, 499--504.

\bibitem{bentley1979}
Bentley, P.~J.
\newblock (1979) The vertebrate urinary bladder: osmoregulatory and other uses.
\newblock {\em The Yale Journal of Biology and Medicine} {\bf 52}, 563.

\bibitem{rao2003}
Rao, S.~G, Walter, J.~S, Jamnia, A, Wheeler, J.~S,  \& Damaser, M.~S.
\newblock (2003) Predicting urethral area from video-urodynamics in women with
  voiding dysfunction.
\newblock {\em Neurourology and Urodynamics} {\bf 22}, 277--283.

\bibitem{walter1993}
Walter, J.~S, Wheeler, J.~S, Morgan, C,  \& Plishka, M.
\newblock (1993) Urodynamic evaluation of urethral opening area in females with
  stress incontinence.
\newblock {\em International Urogynecology Journal} {\bf 4}, 335--341.

\bibitem{barnea2001}
Barnea, O \& Gillon, G.
\newblock (2001) Model-based estimation of male urethral resistance and
  elasticity using pressure--flow data.
\newblock {\em Computers in Biology and Medicine} {\bf 31}, 27--40.

\bibitem{martin2009}
Martin, J.~A \& Hillman, S.~S.
\newblock (2009) The physical movement of urine from the kidneys to the urinary
  bladder and bladder compliance in two anurans.
\newblock {\em Physiological and Biochemical Zoology} {\bf 82}, 163--169.

\bibitem{dass2001}
Dass, N, Mcmurry, G, Greenland, J.~E,  \& Brading, A.~F.
\newblock (2001) Morphological aspects of the female pig bladder neck and
  urethra: quantitative analysis using computer assisted 3-dimensional
  reconstructions.
\newblock {\em The Journal of Urology} {\bf 165}, 1294--1299.

\bibitem{praud2003}
Praud, C, Sebe, P, Mondet, F,  \& Sebille, A.
\newblock (2003) The striated urethral sphincter in female rats.
\newblock {\em Anatomy and Embryology} {\bf 207}, 169--175.

\bibitem{wheeler2012}
Wheeler, A.~P, Morad, S, Buchholz, N,  \& Knight, M.~M.
\newblock (2012) The shape of the urine stream---from biophysics to
  diagnostics.
\newblock {\em PloS One} {\bf 7}, e47133.

\bibitem{kim2012}
Kim, W \& Bush, J.~W.
\newblock (2012) Natural drinking strategies.
\newblock {\em Journal of Fluid Mechanics} {\bf 705}, 7--25.

\bibitem{ogata1970}
Ogata, M, Tomokuni, K,  \& Takatsuka, Y.
\newblock (1970) Urinary excretion of hippuric acid and {\it m}- or {\it
  p}-methylhippuric acid in the urine of persons exposed to vapours of toluene
  and {\it m}- or {\it p}-xylene as a test of exposure.
\newblock {\em British Journal of Industrial Medicine} {\bf 27}, 43--50.

\bibitem{bush2007}
Bush, J.~W, Hu, D.~L,  \& Prakash, M.
\newblock (2007) The integument of water-walking arthropods: form and function.
\newblock {\em Advances in Insect Physiology} {\bf 34}, 117--192.

\bibitem{moore1986}
Moore, C.~L \& Chadwick-Dias, A.-M.
\newblock (1986) Behavioral responses of infant rats to maternal licking:
  Variations with age and sex.
\newblock {\em Developmental Psychobiology} {\bf 19}, 427--438.

\bibitem{prasad2005}
Prasad, S.~R, Menias, C.~O, Narra, V.~R, Middleton, W.~D, Mukundan, G, Samadi,
  N, Heiken, J.~P,  \& Siegel, C.~L.
\newblock (2005) Cross-sectional imaging of the female urethra: Technique and
  results.
\newblock {\em Radiographics} {\bf 25}, 749--761.

\bibitem{cow}
Vyi, M.
\newblock (2005) {J}ersey cattle in {J}ersey
  (\url{http://commons.wikimedia.org/wiki/File:Jersey_cattle_in_Jersey.jpg}).

\bibitem{elephant}
Stolz, G.~M.
\newblock (retrieved at 2013) African {E}lephant
  (\url{http://www.public-domain-image.com/fauna-animals-public-domain-images-pictures/elephant-public-domain-images-pictures/african-elephant-high-resolution.jpg.html}).

\bibitem{bat}
(retrieved at 2013) (\url{http://www.publicdomainpictures.net/}).

\bibitem{nowak1999}
Nowak, R.~M \& Paradiso, J.~L.
\newblock (1999) {\em Walker's mammals of the world.}
\newblock (The Johns Hopkins University Press, Baltimore) Vol.{}~1, 6 edition.

\bibitem{mattern2000}
Mattern, M.~Y \& McLennan, D.~A.
\newblock (2000) Phylogeny and speciation of felids.
\newblock {\em Cladistics} {\bf 16}, 232--253.

\bibitem{wilcox1997}
Wilcox, C.
\newblock (1997) {\em The Great Dane}.
\newblock (Capstone).

\bibitem{wilson2001}
Wilson, D \& Burnie, D.
\newblock (2001) {\em Animal: the definitive visual guide to the world's
  wildlife}.
\newblock (New York: DK Publishing).

\bibitem{miller1997}
Miller-Schroeder, P.
\newblock (1997) {\em Gorillas}, 64.
\newblock (Raintree Steck-Vaughn).

\bibitem{brown1996}
Brown, G.
\newblock (1996) {\em The Great Bear Almanac}.
\newblock (Globe Pequot).

\bibitem{starkey1992}
Starkey, P, Mwenya, E,  \& Stares, J.
\newblock (1992) {\em Improving animal traction technology}.
\newblock pp. 18--23.

\bibitem{lynette2013}
Lynette, R.
\newblock (2013) {\em South American Tapirs}.
\newblock (Bearport Publishing).

\bibitem{linnaeus1758}
Linnaeus, C.
\newblock (1758) {\em Systema naturae per regna tria naturae, secundum classes,
  ordines, genera, species, cum characteribus, differentiis, synonymis, locis}.
\newblock (Laurentii Salvii, Stockholm), 10 edition.

\bibitem{kingdon1988}
Kingdon, J.
\newblock (1988) {\em East African Mammals: An Atlas of Evolution in Africa,
  Part A: Carnivores}.
\newblock (University of Chicago Press) Vol.{}~3.

\bibitem{marvin1992}
Hall, M.~H \& Comerford, P.~M.
\newblock (1992) Pasture and hay for horses.
\newblock {\em Cooperative Extension, The Pennsylvania State University} p.~32.

\bibitem{bongianni1988}
Bongianni, M.
\newblock (1988) {\em Simon and Schuster's guide to horses and ponies of the
  world}.
\newblock (Simon and Schuster).

\bibitem{potts1997}
Potts, S.
\newblock (1997) {\em The American Bison}.
\newblock (Capstone).

\bibitem{toon2002}
Toon, A \& Toon, S.
\newblock (2002) {\em Rhinos}.
\newblock (Voyageur Press).

\bibitem{shoshani1982}
Shoshani, J \& Eisenberg, J.~F.
\newblock (1982) Elephas maximus.
\newblock {\em Mammalian Species}.

\bibitem{clair1999}
Clair, M.~B, Sowers, A.~L, Davis, J.~A,  \& Rhodes, A.~L.
\newblock (1999) Urinary bladder catheterization of female mice and rats.
\newblock {\em Journal of the American Association for Laboratory Animal
  Science} {\bf 38}, 78--79.

\bibitem{chang2000}
Chang, S.-C, Chern, I,  \& Bown, S.
\newblock (2000) Photodynamic therapy of rat bladder and urethra: evaluation of
  urinary and reproductive function after inducing protoporphyrin ix with
  5-aminolaevulinic acid.
\newblock {\em BJU International} {\bf 85}, 747--753.

\bibitem{kamo2004}
Kamo, I, Cannon, T.~W, Conway, D.~A, Torimoto, K, Chancellor, M.~B, de~Groat,
  W.~C,  \& Yoshimura, N.
\newblock (2004) The role of bladder-to-urethral reflexes in urinary continence
  mechanisms in rats.
\newblock {\em American Journal of Physiology-Renal Physiology} {\bf 287},
  F434--F441.

\bibitem{vogel2007}
Vogel, H.
\newblock (2007) {\em Drug Discovery and Evaluation: Pharmacological Assays}.
\newblock (Springer), 3 edition.

\bibitem{johnston1985}
Johnston, G, Osborne, C,  \& Jessen, C.
\newblock (1985) Effects of urinary bladder distension on the length of the dog
  and cat urethra.
\newblock {\em American Journal of Veterinary Research} {\bf 46}, 509.

\bibitem{kropp1998}
Kropp, B.~P, Ludlow, J.~K, Spicer, D, Rippy, M.~K, Badylak, S.~F, Adams, M.~C,
  Keating, M.~A, Rink, R.~C, Birhle, R,  \& Thor, K.~B.
\newblock (1998) Rabbit urethral regeneration using small intestinal submucosa
  onlay grafts.
\newblock {\em Urology} {\bf 52}, 138--142.

\bibitem{takeda1995}
Takeda, M \& Lepor, H.
\newblock (1995) Nitric oxide synthase in dog urethra: a histochemical and
  pharmacological analysis.
\newblock {\em British Journal of Pharmacology} {\bf 116}, 2517--2523.

\bibitem{kohler2008}
Kohler, T, Yadven, M, Manvar, A, Liu, N,  \& Monga, M.
\newblock (2008) The length of the male urethra.
\newblock {\em International Braz J Urol} {\bf 34}, 451--456.

\bibitem{lueders2012}
Lueders, I, Luther, I, Scheepers, G,  \& van~der Horst, G.
\newblock (2012) Improved semen collection method for wild felids: urethral
  catheterization yields high sperm quality in african lions (\emph{Panthera
  leo}).
\newblock {\em Theriogenology}.

\bibitem{hildebrandt2000}
Hildebrandt, T.~B, G{\"o}ritz, F, Pratt, N.~C, Brown, J.~L, Montali, R.~J,
  Schmitt, D.~L, Fritsch, G,  \& Hermes, R.
\newblock (2000) Ultrasonography of the urogenital tract in elephants
  (\emph{Loxodonta africana} and \emph{Elephas maximus}): an important tool for
  assessing female reproductive function.
\newblock {\em Zoo Biology} {\bf 19}, 321--332.

\bibitem{fowler2006}
Fowler, M.~E \& Mikota, S.~K.
\newblock (2006) {\em Biology, Medicine, and Surgery of Elephants}.
\newblock (Wiley-Blackwell).

\bibitem{krumrey1968}
Krumrey, W.~A \& Buss, I.~O.
\newblock (1968) Age estimation, growth, and relationships between body
  dimensions of the female african elephant.
\newblock {\em Journal of Mammalogy} pp. 22--31.

\bibitem{de2008}
de~Souza, A. B.~G, Suaid, H.~J, Suaid, C.~A, Jr, S.~T, Cologna, A.~J,  \&
  Martins, A. C.~P.
\newblock (2008) Comparison of two experimental models of urodynamic evaluation
  in female rats.
\newblock {\em Acta Cirurgica Brasileira} {\bf 23}.

\bibitem{russell1996}
Russell, B, Baumann, M, Heidkamp, M,  \& Svanborg, A.
\newblock (1996) Morphometry of the aging female rat urethra.
\newblock {\em International Urogynecology Journal} {\bf 7}, 30--36.

\bibitem{kunstyvr1982}
Kunst{\`y}{\v{r}}, I, K{\"u}pper, W, Weisser, H, Naumann, S,  \& Messow, C.
\newblock (1982) Urethral plug-a new secondary male sex characteristic in rat
  and other rodents.
\newblock {\em Laboratory Animals} {\bf 16}, 151--155.

\bibitem{root1996}
Root, M.~V, Johnston, S.~D, Johnston, G.~R,  \& Olson, P.~N.
\newblock (1996) The effect of prepuberal and postpuberal gonadectomy on penile
  extrusion and urethral diameter in the domestic cat.
\newblock {\em Veterinary Radiology \& Ultrasound} {\bf 37}, 363--366.

\bibitem{wang2010}
Wang, Y, Nagarajan, U, Hennings, L, Bowlin, A.~K,  \& Rank, R.~G.
\newblock (2010) Local host response to chlamydial urethral infection in male
  guinea pigs.
\newblock {\em Infection and Immunity} {\bf 78}, 1670--1681.

\bibitem{gray1918}
Gray, H.
\newblock (1918) {\em Anatomy of the Human Body}.
\newblock (Lea \& Febiger).

\bibitem{tsujimoto2003}
Tsujimoto, Y, Nose, Y,  \& Ohba, K.
\newblock (2003) Experimental and clinical trial of measuring urinary velocity
  with the pitot tube and a transrectal ultrasound guided video urodynamic
  system.
\newblock {\em International Journal of Urology} {\bf 10}, 30--35.

\bibitem{pozor2002}
Pozor, M \& McDonnell, S.
\newblock (2002) Ultrasonographic measurements of accessory sex glands,
  ampullae, and urethra of normal stallions of various size types.
\newblock {\em Theriogenology} {\bf 58}, 1425--1433.

\bibitem{hildebrandt1998}
Hildebrandt, T.~B, G{\"o}ritz, F, Pratt, N.~C, Schmitt, D.~L, Quandt, S, Raath,
  J,  \& Hofmann, R.~R.
\newblock (1998) Reproductive assessment of male elephants (\emph{Loxodonta
  africana} and \emph{Elephas maximus}) by ultrasonography.
\newblock {\em Journal of Zoo and Wildlife Medicine} pp. 114--128.

\bibitem{bailey1975}
Bailey, C.
\newblock (1975) Siliceous urinary calculi in bulls, steers, and partial
  castrates.
\newblock {\em Canadian Journal of Animal Science} {\bf 55}, 187--191.

\bibitem{tasaki2009}
Tasaki, M, Umemura, T, Kijima, A, Inoue, T, Okamura, T, Kuroiwa, Y, Ishii, Y,
  \& Nishikawa, A.
\newblock (2009) Simultaneous induction of non-neoplastic and neoplastic
  lesions with highly proliferative hepatocytes following dietary exposure of
  rats to tocotrienol for 2 years.
\newblock {\em Archives of Toxicology} {\bf 83}, 1021--1030.

\bibitem{perrin2003}
Perrin, D, Soulage, C, Pequignot, J,  \& Geloen, A.
\newblock (2003) Resistance to obesity in lou/c rats prevents ageing-associated
  metabolic alterations.
\newblock {\em Diabetologia} {\bf 46}, 1489--1496.

\bibitem{sturman1985}
Sturman, J, Moretz, R, French, J,  \& Wisniewski, H.
\newblock (1985) Postnatal taurine deficiency in the kitten results in a
  persistence of the cerebellar external granule cell layer: correction by
  taurine feeding.
\newblock {\em Journal of Neuroscience Research} {\bf 13}, 521--528.

\bibitem{scott1970}
Scott, P \& Hafez, E.
\newblock (1970) {\em Reproduction and Breeding Techniques for Laboratory
  Animals}.
\newblock (Lea \& Febiger).

\bibitem{lein1983}
Lein, D \& Concannon, P.
\newblock (1983) Infertility and fertility treatments and management in the
  queen and tomcat.
\newblock {\em Current Therapy VIII. Kirk, R.(ed). WB Saunders Co.
  Philadelphia} pp. 936--987.

\bibitem{ogden2004}
Ogden, C.~L, Fryar, C.~D, Carroll, M.~D,  \& Flegal, K.~M.
\newblock (2004) {\em Mean Body Weight, Height, and Body Mass Index: United
  States 1960-2002}.
\newblock (Department of Health and Human Services, Centers for Disease Control
  and Prevention, National Center for Health Statistics).

\bibitem{treuting2011}
Treuting, P \& Dintzis, S.~M.
\newblock (2011) {\em Comparative Anatomy and Histology: A mouse and Human
  Atlas}.
\newblock (Academic Press).

\bibitem{caceci}
Caceci, T.
\newblock (year?) Canine penis; h \& e stain, paraffin section (decalcified),
  20x
  (\url{http://www.vetmed.vt.edu/education/curriculum/vm8304/lab_companion/histo-path/vm8054/labs/Lab23/EXAMPLES/EXURETH.HTM}).

\bibitem{skarva}
Skarva, F.
\newblock (year?) Cross-section of a normal human penis showing the urethra and
  corpora spongiosum
  (\url{http://www.allposters.com/-sp/Cross-Section-of-a-Normal-Human-Penis-Showing-the-Urethra-and-Corpora-Spongiosum-H-E-Stain-LM-X12-Posters_i9005304_.htm}).

\bibitem{pandita2000}
Pandita, R.~K, Fujiwara, M, Alm, P,  \& Andersson, K.-E.
\newblock (2000) Cystometric evaluation of bladder function in non-anesthetized
  mice with and without bladder outlet obstruction.
\newblock {\em The Journal of Urology} {\bf 164}, 1385--1389.

\bibitem{birder2002}
Birder, L, Nakamura, Y, Kiss, S, Nealen, M, Barrick, S, Kanai, A, Wang, E,
  Ruiz, G, De~Groat, W, Apodaca, G,  et~al.
\newblock (2002) Altered urinary bladder function in mice lacking the vanilloid
  receptor trpv1.
\newblock {\em Nature Neuroscience} {\bf 5}, 856--860.

\bibitem{herrera2010}
Herrera, G.~M \& Meredith, A.~L.
\newblock (2010) Diurnal variation in urodynamics of rat.
\newblock {\em PloS One} {\bf 5}, e12298.

\bibitem{thor1995}
Thor, K.~B \& Katofiasc, M.~A.
\newblock (1995) Effects of duloxetine, a combined serotonin and norepinephrine
  reuptake inhibitor, on central neural control of lower urinary tract function
  in the chloralose-anesthetized female cat.
\newblock {\em Journal of Pharmacology and Experimental Therapeutics} {\bf
  274}, 1014--1024.

\bibitem{abdel2001}
Abdel-Gawad, M, Boyer, S, Sawan, M,  \& Elhilali, M.
\newblock (2001) Reduction of bladder outlet resistance by selective
  stimulation of the ventral sacral root using high frequency blockade: a
  chronic study in spinal cord transected dogs.
\newblock {\em The Journal of Urology} {\bf 166}, 728--733.

\bibitem{atalan1998}
Atalan, G, Barr, F.~J,  \& Holt, P.~E.
\newblock (1998) Estimation of bladder volume using ultrasonographic
  determination of cross-sectional areas and linear measurements.
\newblock {\em Veterinary Radiology \& Ultrasound} {\bf 39}, 446--450.

\bibitem{higgins2006}
Higgins, A.~J \& Snyder, J.~R.
\newblock (2006) {\em The Equine Manual}.
\newblock (Elsevier Saunders).

\bibitem{ishizuka1996}
Ishizuka, O, Persson, K, Mattiasson, A, Naylor, A, Wyllie, M,  \& Andersson,
  K.-E.
\newblock (1996) Micturition in conscious rats with and without bladder outlet
  obstruction: role of spinal $\alpha$1-adrenoceptors.
\newblock {\em British Journal of Pharmacology} {\bf 117}, 962--966.

\bibitem{hinman1971}
Hinman, F.
\newblock (1971) {\em Hydrodynamics of Micturition}.
\newblock (Thomas).

\bibitem{van1995}
Van~Asselt, E, Groen, J,  \& Van~Mastrigt, R.
\newblock (1995) A comparative study of voiding in rat and guinea pig:
  simultaneous measurement of flow rate and pressure.
\newblock {\em American Journal of Physiology-Regulatory, Integrative and
  Comparative Physiology} {\bf 269}, R98--R103.

\bibitem{schmidt2003}
Schmidt, F, Yoshimura, Y, Shin, P.~Y,  \& Constantinou, C.~E.
\newblock (2003) Comparative urodynamic patterns of bladder pressure,
  contractility and urine flow in man and rat during micturition.
\newblock {\em Acta Pathologica Microbiologica et Immunologica Scandinavica -
  Supplementum} {\bf 111}, 39--44.

\bibitem{masumori1996}
Masumori, N, Tsukamoto, T, Kumamoto, Y, Miyake, H, Rhodes, T, Girman, C.~J,
  Guess, H.~A, Jacobsen, S.~J,  \& Lieber, M.~M.
\newblock (1996) Japanese men have smaller prostate volumes but comparable
  urinary flow rates relative to american men: results of community based
  studies in 2 countries.
\newblock {\em The Journal of Urology} {\bf 155}, 1324--1327.

\bibitem{folkestad2004}
Folkestad, B \& Sp{\aa}ngberg, A.
\newblock (2004) Timed micturition and maximum urinary flow rate in randomly
  selected symptom-free males.
\newblock {\em Scandinavian Journal of Urology and Nephrology} {\bf 38},
  136--142.

\bibitem{schmidt2002}
Schmidt, F, Shin, P, Jorgensen, T.~M, Djurhuus, J.~C,  \& Constantinou, C.~E.
\newblock (2002) Urodynamic patterns of normal male micturition: influence of
  water consumption on urine production and detrusor function.
\newblock {\em The Journal of Urology} {\bf 168}, 1458--1463.

\end{thebibliography}

 \clearpage
      \begin{table*}
           \begin{center}
                \includegraphics[width = 6.5in,keepaspectratio=true]
             {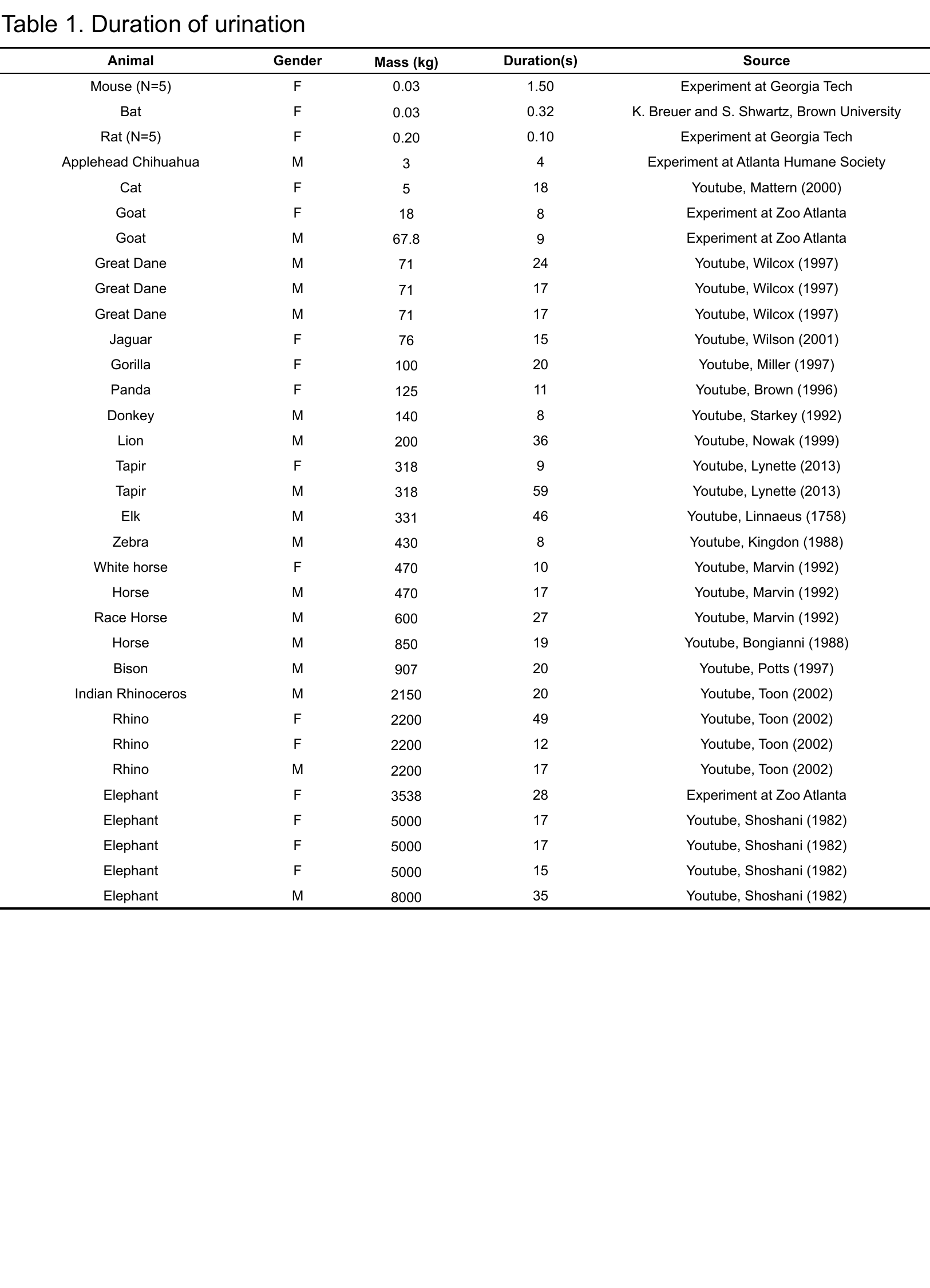}
         \end{center}
    \end{table*} 
    
           \begin{table*}
           \begin{center}
                \includegraphics[width = 6.5in,keepaspectratio=true]
             {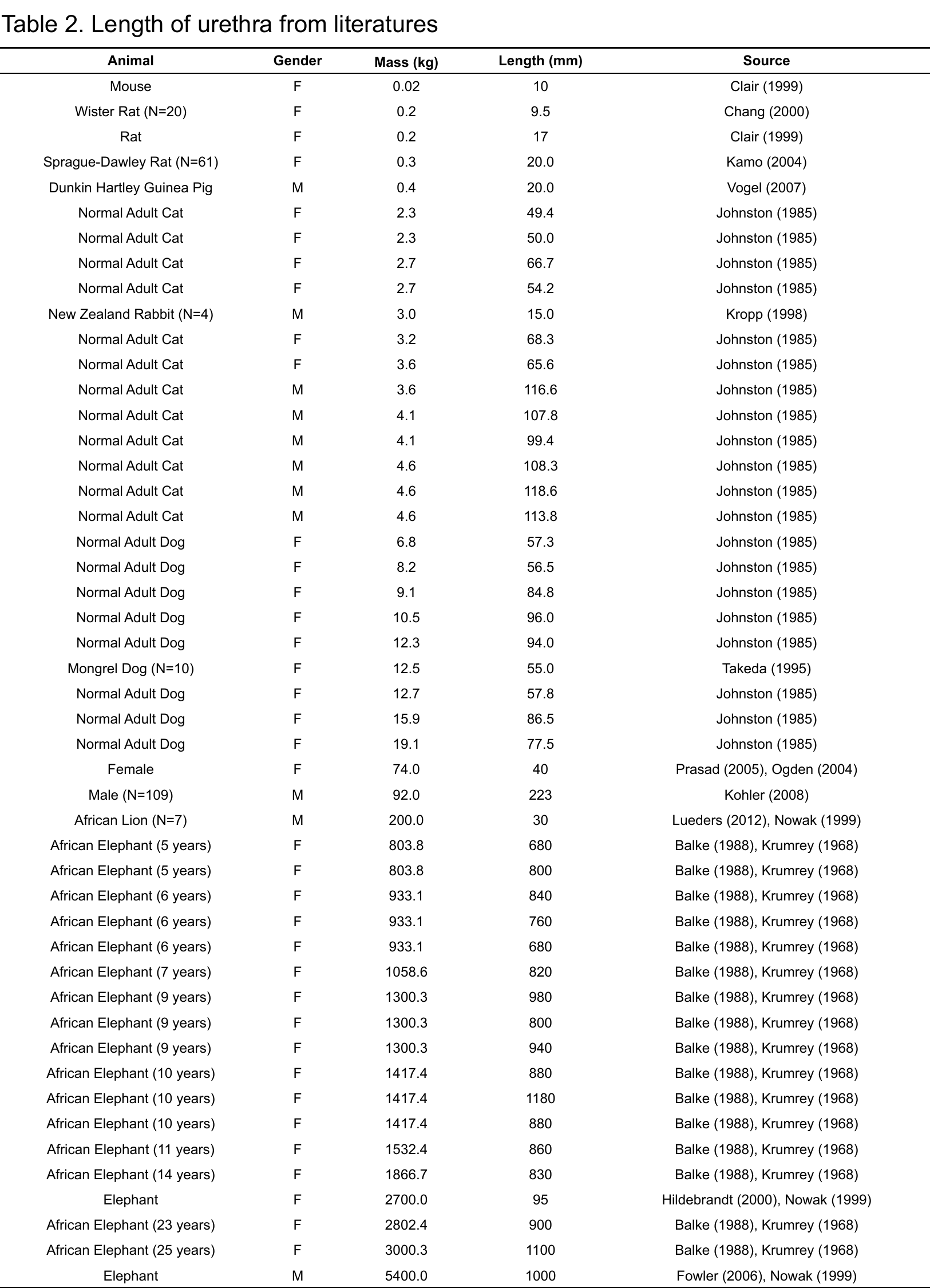}
         \end{center}
    \end{table*} 
    
    
          \begin{table*}
           \begin{center}
                \includegraphics[width = 6.5in,keepaspectratio=true]
             {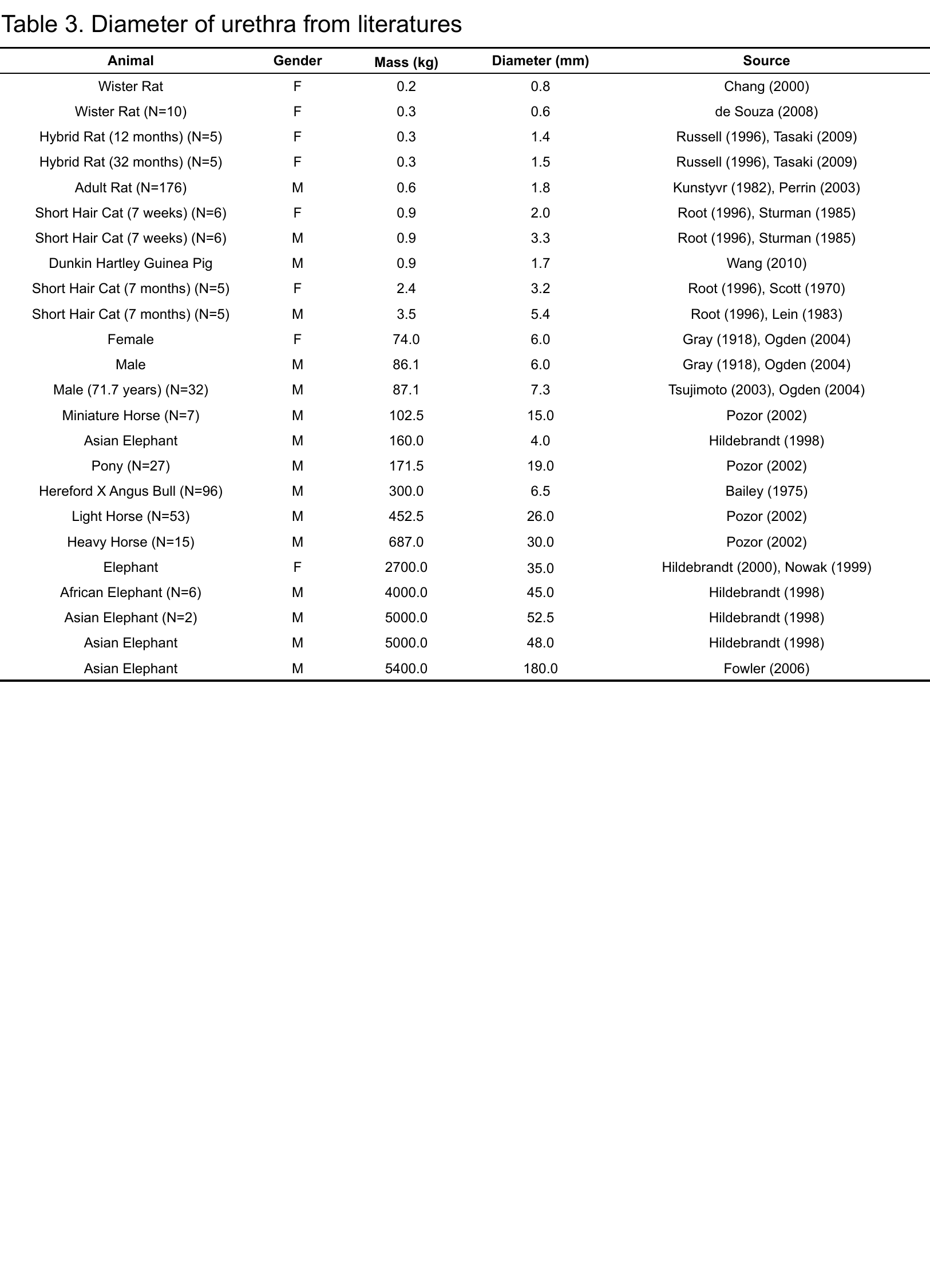}
         \end{center}
    \end{table*} 
    
              \begin{table*}
           \begin{center}
                \includegraphics[width = 6.5in,keepaspectratio=true]
             {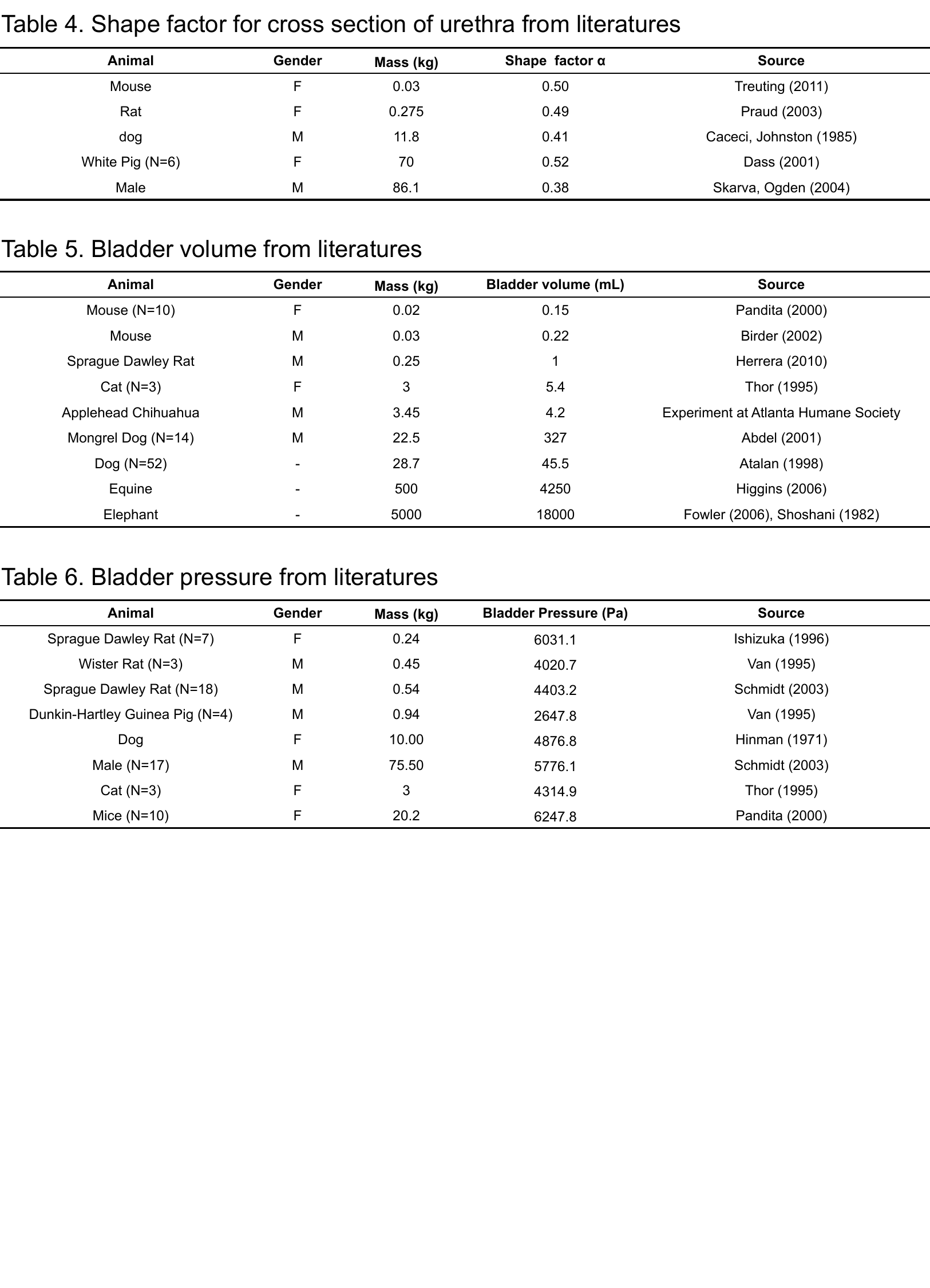}
         \end{center}
    \end{table*} 
    
            \begin{table*}
           \begin{center}
                \includegraphics[width = 6.5in,keepaspectratio=true]
             {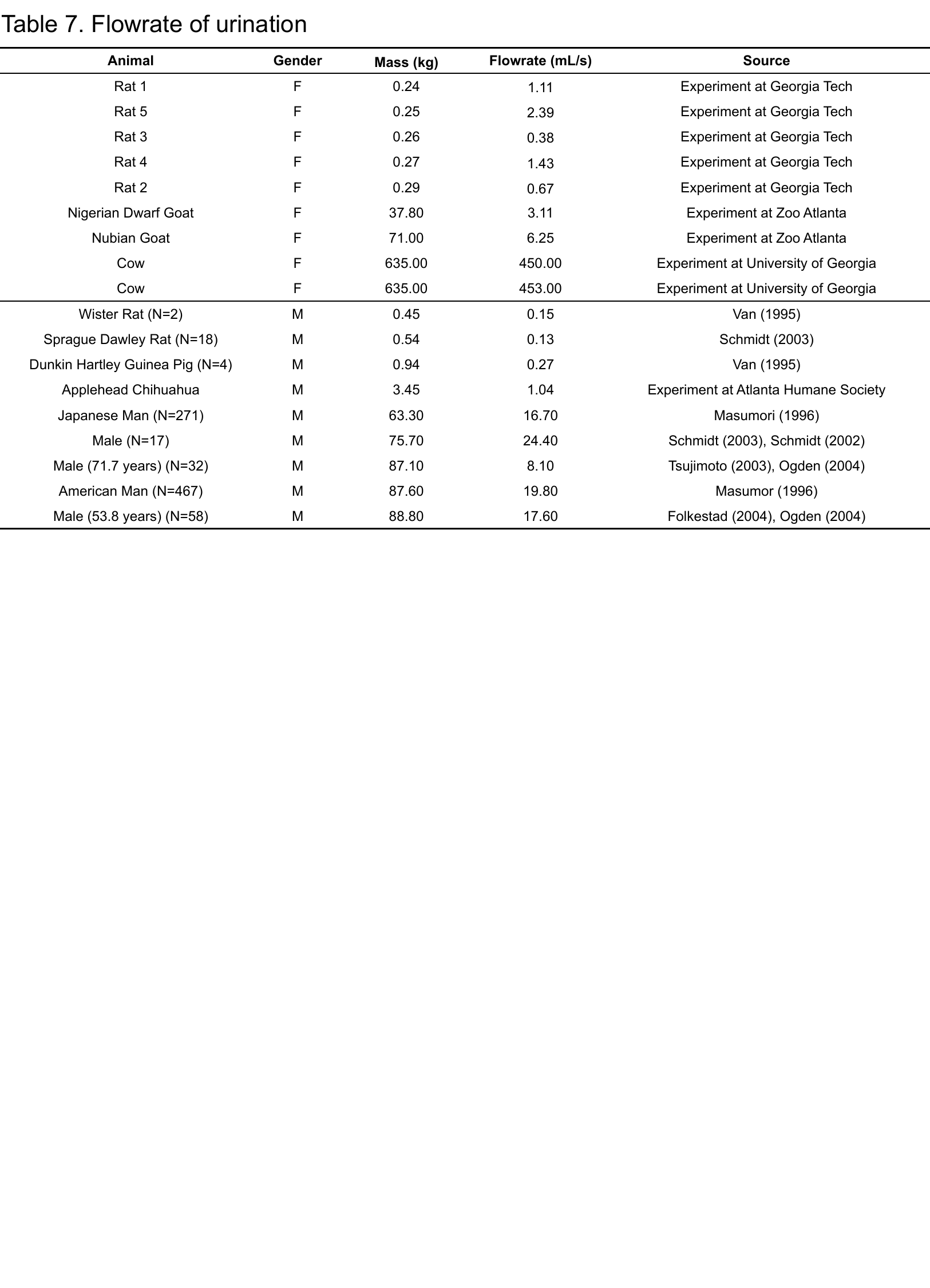}
         \end{center}
    \end{table*}

      \begin{figure*}
       \begin{center}
                \includegraphics[width = 7in,keepaspectratio=true]
             {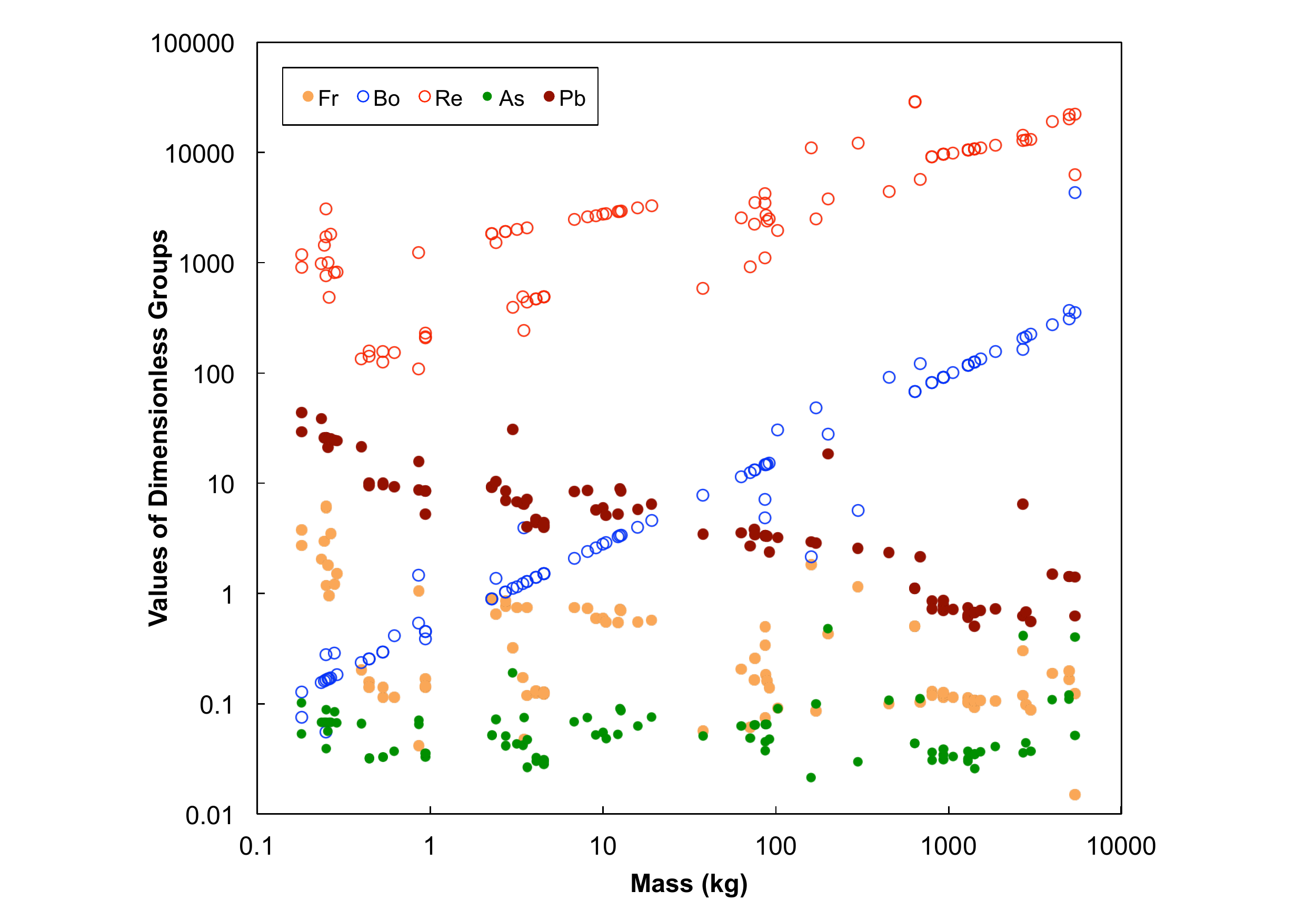}
         \end{center}
      \caption{The magnitude of five dimensionless group, the Froud $Fr$, Bond $Bo$, Reynolds $Re$, aspect ratio $As$, and ratio of bladder and gravity pressure $Pb$.}
      \label{dimensionless}
    \end{figure*}
    
  \end{document}